\documentclass{article}

\usepackage{microtype}
\usepackage{graphicx}
\usepackage{booktabs}
\usepackage{subcaption}
\usepackage[sort&compress]{natbib}
\usepackage{amsmath,amssymb,amsfonts}
\usepackage{algorithmic}
\usepackage{textcomp}
\usepackage{xcolor}
\usepackage{colortbl}
\usepackage{ctable}
\usepackage{float}
\usepackage[colorlinks=true, linkcolor=black, citecolor=black, urlcolor=blue]{hyperref}

\usepackage[accepted]{preprint2col}

\pptitlerunning{Muting Tokens: Token Sparsification for Efficient Audio Transformers}

\begin{document}

\twocolumn[
\pptitle{Muting Tokens: Exploring Token Sparsification\\ for Efficient Audio Transformers}

\ppsetsymbol{equal}{\dag}

\begin{ppauthorlist}
\ppauthor{Shreya Sajid}{tue}
\ppauthor{Fabrizio Julian Piva}{equal,nxp}
\ppauthor{Hiram Rayo Torres Rodriguez}{equal,nxp}
\ppauthor{Elif Ceren Gok Yildirim}{tue}
\ppauthor{Murat Onur Yildirim}{tue}
\ppauthor{Joaquin Vanschoren}{tue}
\end{ppauthorlist}

\begin{center}
\small\texttt{s.sajid@tue.nl, fabriziojpiva@gmail.com, hiram.rayotorresrodriguez@nxp.com,}\\
\small\texttt{e.c.gok@tue.nl, m.o.yildirim@tue.nl, j.vanschoren@tue.nl}
\end{center}

\ppaffiliation{tue}{Eindhoven University of Technology, Eindhoven, Netherlands}
\ppaffiliation{nxp}{NXP Semiconductors, Eindhoven, Netherlands}

\ppkeywords{Token Sparsification, Audio Transformers, Efficient Inference}

\vskip 0.3in

\begin{abstract}
Transformer-based models have shown strong performance across diverse tasks, but their reliance on self-attention mechanisms often leads to high computational costs. Token sparsification, which reduces the number of tokens processed in transformer layers, offers a promising solution to maintain performance while improving efficiency. However, although widely studied in the vision and language domains, token sparsification remains underexplored in audio, with existing work often limited to a single method, model size, or dataset. This work presents a comprehensive evaluation of token sparsification techniques for audio transformers. We assess a range of token pruning and merging strategies across multiple datasets and model sizes, under both post-training and fine-tuning settings. Furthermore, we introduce a novel sparsification approach that combines attention scores with spectral cues derived from the frequency content of token embeddings. Our experiments show that frequency-aware approaches, particularly those based on spectral flatness and low-frequency energy, consistently outperform attention-only methods. We also find that sparsification based on spectrogram patch intensity is especially effective for longer audio inputs. These results highlight the value of incorporating frequency-domain and input-specific information into sparsification strategies to enhance performance in audio tasks.
\end{abstract}
]

\printAffiliationsAndNotice{\ppEqualContribution}

\section{Introduction} \label{sec:intro}

Transformer-based models have demonstrated the ability to achieve state-of-the-art performance across various domains such as Natural Language Processing (NLP) \cite{vaswani2017attention, kenton2019bert} and Computer Vision \cite{khan2021transformers, radford2021learning}. However, the self-attention mechanism can be computationally intensive, as its complexity scales quadratically with the number of input tokens \cite{Bolya_Fu_Dai_Zhang_Feichtenhofer_Hoffman_2023, fayyaz2022adaptive}. To address this challenge, recent work has explored compression techniques such as pruning \cite{chavan2022vision, song2022cp}, quantization \cite{li2022q, lin2022fq}, and knowledge distillation (KD) \cite{ren2022civt, rao2021dynamicvit}, in an effort to reduce model size and inference latency while maintaining task performance.

\begin{figure}[t!]
    \centering
    \fbox{\includegraphics[width=\dimexpr 0.88\linewidth-2\fboxsep-2\fboxrule\relax]{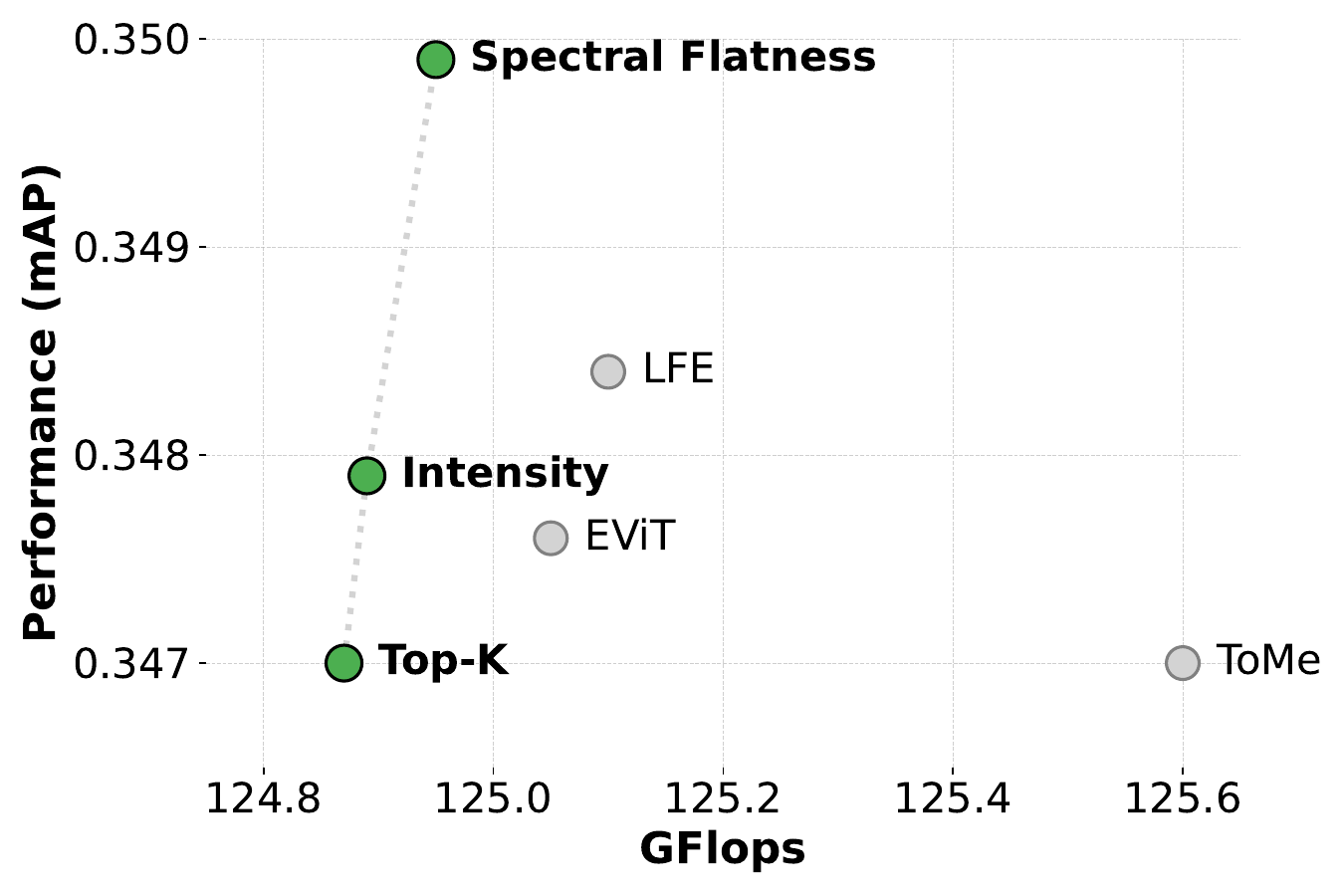}}
    % \fbox{\parbox[b][5cm][c]{8cm}{\centering Figure Placeholder}}
    \vspace{0.5em}
\caption{Post-training sparsification results on the AST-Base \cite{gong2021} model, evaluated on AudioSet-20K \cite{audioset} at a 50\% keep rate. Methods that include frequency-domain cues, such as the proposed \textbf{Spectral Flatness (SF)} and the adapted \textbf{Low-Frequency Energy (LFE) \cite{wang2022vtclfc}}, outperform others in terms of mean average precision (mAP), a standard metric for multi-label classification. The baseline (without sparsification) achieves an average mAP of 0.3572. The Pareto front is highlighted to illustrate the performance-efficiency trade-off. }

    \label{fig:mainPareto}
\end{figure} 

Recently, token sparsification, a transformer-specific optimization method that discards, merges, or refines input tokens, has emerged as a promising alternative to reduce the computational cost of self-attention. For example, the Expediting Vision Transformers (EViT) method achieves up to 50\% faster inference on ImageNet \cite{imagenet}, with only a 1.3 percentage point drop in accuracy when retaining 50\% of the tokens, referred to as the \textit{keep rate}~\cite{evit2022}. 

A growing body of work has explored token sparsification techniques across various tasks, particularly in image classification \cite{ppt2023, ltmp2023, evit2022, cfvit2023, toe2024, kong2022spvit}, video classification \cite{tokenlearner2022, haltingvt2024}, and even large language models \cite{tcra2024, lllingue2023, fastv2024, pumer2023}. Although Vision Transformers (ViT) have demonstrated notable success in audio tasks by using spectrograms as input representations, the application of token sparsification in this domain remains largely underexplored \cite{gong2021, gong2022, gong2022passt, huang2022, chen2022}. This gap is particularly important because spectrogram inputs yield more tokens as the input duration increases, which can make self-attention increasingly expensive for longer audio clips \cite{gong2021}. Addressing this challenge could not only help advance efficient transformers' research, but also enable practical, large-scale deployment of audio-based systems, from everyday voice assistants to real-time transcription services at the edge.

Recent work \cite{lee2025token} has begun to explore the effectiveness of token sparsification in audio tasks. However, while showing promising results, this study presents considerable limitations. First, it restricts its scope to only two sparsification methods, Top-K pruning~\cite{haurum2022tokens} and  Intensity pruning~\cite{lee2025token}, overlooking other promising approaches such as token merging \cite{Bolya_Fu_Dai_Zhang_Feichtenhofer_Hoffman_2023}. Second, their analysis is restricted to only the ViT-Base model~\cite{dosovitskiy2021an} limiting the generalizability of their conclusions to a single model size. Third, the study focuses solely on applying sparsification during fine-tuning, whereas many real-world deployment scenarios favor post-training sparsification methods, mainly due to the widespread availability of pretrained models, and the training costs associated with larger datasets \cite{Ilhan_2024_CVPR, chen2025tunecompjointfinetuningcompression}. Finally, it reports suboptimal performance of the intensity pruning method at low keep rates, without investigating the potential influence of input-dependent factors such as the audio duration.

In this work, we systematically evaluate a broad range of token sparsification methods for audio transformers, examining their performance across multiple model sizes and datasets, under two key settings: fine-tuning and post-training sparsification. To ensure a fair comparison, we conduct all experiments within a unified training and inference pipeline, establishing a consistent benchmark for evaluating sparsification techniques in this domain. We also revisit the intensity pruning method and show that its effectiveness improves when combined with attention scores. Furthermore, we propose a novel sparsification approach that integrates attention and frequency-based cues, drawing on insights from audio signal processing. As illustrated in Fig.~\ref{fig:mainPareto}, methods that incorporate frequency-domain information consistently outperform existing sparsification approaches across our experimental setup. Our contributions can be summarized as:
\begin{itemize}
    \item We conduct a comprehensive evaluation of token sparsification methods for audio transformers, including token pruning and merging approaches under a unified evaluation setup.
    \item We demonstrate the effectiveness of intensity-based pruning, showing that when combined with attention scores, it can outperform pure attention methods for audio inputs longer than 2 seconds across our experimental setup.
    \item We introduce a novel sparsification method based on spectral flatness, and show that it can consistently improve performance across a wide range of model sizes and datasets.
\end{itemize}

\section{Related Work}

\subsection{Token Sparsification: Overview} 
\label{sec:related_work_overview}
Token sparsification reduces the number of tokens processed within transformer layers, with existing methods broadly falling into two main categories \cite{haurum2022tokens}. \textbf{\textit{Token pruning}} methods \cite{dou2022tore,fayyaz2022adaptive,kong2022spvit,li2022sait,liang2022evit,long2022beyond,pan2021iared,tang2022patch,xu2022evovit,yin2022avit} reduce the number of tokens by completely removing tokens, at the risk of discarding useful information. In contrast, \textbf{\textit{token merging}} methods \cite{Bolya_Fu_Dai_Zhang_Feichtenhofer_Hoffman_2023,haurum2022multi,marin2023token,renggli2022learning,ryoo2021tokenlearner,wu2021centroid,xu2022groupvit,zeng2022not,zong2022selfslimmed} combine multiple tokens into a single token or cluster, but may introduce noise through averaging.

Our study includes representative methods from both sparsification categories to ensure diversity in evaluation, addressing the lack of systematic assessment in audio tasks.

\subsection{Token Sparsification Methods for Audio}  
\label{sec:related_work_token_sparsification_for_audio}
Recent efforts to apply token sparsification to audio tasks have primarily adapted techniques originally developed for vision models. Applied on a post-training manner, Token Merging (ToMe) \cite{Bolya_Fu_Dai_Zhang_Feichtenhofer_Hoffman_2023} merges similar tokens to reduce the input size, achieving substantial throughput gains with merely a 0.4 percentage point drop in mAP on AudioSet-2M~\cite{audioset}. However, its evaluation is limited to a single dataset and a masked autoencoding (MAE) framework \cite{DBLP:journals/corr/abs-2111-06377, huang2022}, which is inherently designed to operate on sparse inputs. This may reduce the model’s sensitivity to sparsification, making it unclear how ToMe performs in settings that rely on full token visibility. Without evaluation on non-MAE models, its general applicability remains uncertain, an aspect we address in this work.

A different line of work explores sparsification in audio transformers during training. PaSST \cite{gong2022passt} employs “patchout” which randomly drops a subset of spectrogram patches to reduce computation and act as a regularizer. While effective, these approaches require training with patchout, limiting their practicality. In contrast, post-training sparsification offers a broadly applicable and cost-efficient alternative, especially when retraining is costly or infeasible.

\subsection{Studies on Token Sparsification for Audio}
Beyond these early efforts, recent work has begun to explore sparsification methods specifically for audio. As mentioned in Sec.~\ref{sec:intro}, the authors in \cite{lee2025token} investigate Top-K pruning and Intensity pruning. While this represents a promising direction, the study remains narrow in scope, focusing on just two methods, a single ViT-Base configuration, and exclusively on fine-tuning scenarios. This highlights the need for broader studies that evaluate different sparsification approaches, model sizes, and across both fine-tuning and post-training contexts, a gap our work seeks to address.

\section{Methodology}
\label{sec:methodology}
In this study, we pursue two complementary directions. First, we evaluate established token sparsification methods developed for vision, in audio tasks, providing the first systematic comparison of their behavior across audio benchmarks. Second, we propose lightweight, task-specific token importance estimation strategies that incorporate audio-relevant cues such as spectrogram intensity and frequency-domain features.

\subsection{Token Sparsification Methods}
\label{sec:baselines}

We evaluate three sparsification methods: Top-K \cite{haurum2022tokens}, Expediting Vision Transformers (EViT) \cite{evit2022} and ToMe \cite{Bolya_Fu_Dai_Zhang_Feichtenhofer_Hoffman_2023}. These methods were selected based on their strong performance in vision tasks \cite{haurum2022tokens} and their representation of the two main paradigms in token sparsification: \emph{pruning} (Top-K and EViT) and \emph{merging} (ToMe). Note that while some of these methods have been applied to audio in a limited setting, as mentioned in Sec.~\ref{sec:related_work_token_sparsification_for_audio}, they have not yet been systematically compared or evaluated across benchmarks. \textbf{Top-K} is a widely used baseline, where the attention between spatial tokens and the class (CLS) token are used to rank token importance. At each reduction stage $s$, the $K_s$ most attended tokens are kept. \textbf{EViT} extends this approach by introducing a fused token that preserves information from discarded tokens. At each stage $s$, tokens not in the Top-K set are averaged and weighted by their CLS attention scores. \textbf{ToMe} reduces tokens by merging similar token pairs. The token set is split into two partitions, and each token is matched to its most similar counterpart based on cosine similarity. The top matching pairs are then merged by averaging their features.

\begin{figure}[t]
\centering
    \vspace{0.9em}
  \includegraphics[width = 1.0\linewidth]{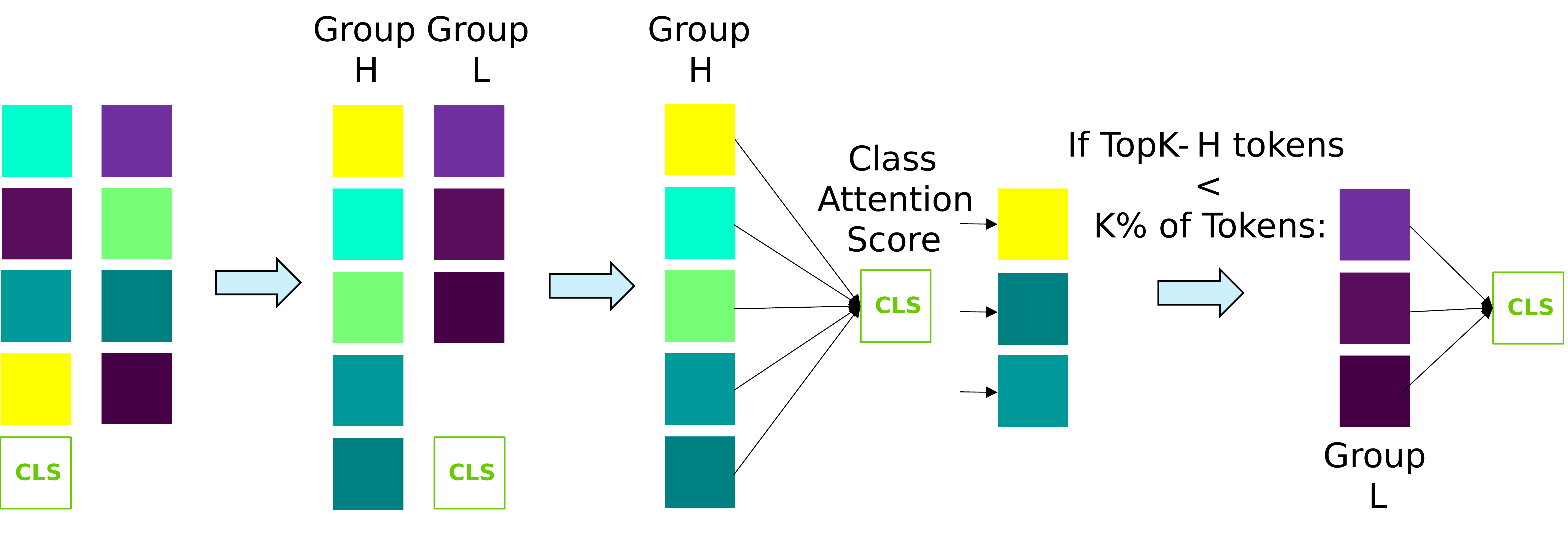}
  \vspace{0.3em}
  \caption{Intensity-guided token pruning. Tokens are divided into high- (\( H \)) and low- (\( L \)) intensity groups based on an applied threshold. The high-intensity set is then ranked using CLS attention scores, and the Top-\( K \) tokens are selected. If fewer than \( K \) tokens are available in \( H \), the remaining slots are filled from the low-intensity set.}
  \label{fig:intensity}
  \vspace{-1em}
\end{figure}

\subsection{Task-specific Token Importance Estimation Strategies}
\label{sec:impstrats}
Most sparsification methods rely on internal model signals, such as attention weights or intermediate representations like attention keys. However, in post-training settings where model parameters remain fixed, we hypothesize that lightweight, task-specific heuristics, can more effectively identify informative tokens, leading to improved performance under sparsification. In audio tasks, spectrograms encode energy across time and frequency, but these patterns may not be fully captured by attention alone~\cite{ristea2022septrseparabletransformeraudio}. Prior work highlights the utility of frequency-domain processing in transformers~\cite{tamkin2020prism, lee2021fnet, rao2021gfnet, patro2025spectformer}, and that structure may not always be visible in raw spectrograms~\cite{ezzat2007spectro, atlas2003modulation}. Motivated by this, we explore three lightweight heuristics for token importance, one input patch intensity-based, and two frequency-based, that can be combined with many existing methods: 1) intensity, 2) low-frequency energy \cite{wang2022vtclfc} and 3) spectral flatness. \\

\noindent \textbf{Intensity.} Following~\cite{lee2025token}, we compute the mean log-Mel energy of each spectrogram patch as a lightweight salience proxy, based on the hypothesis that high intensity regions reflect stronger signal content. However, using intensity alone can be unreliable, as it can sometimes highlight background noise or non-discriminative events with high energy \cite{Bhattacharjee2022FBAS, lee2025token}. We therefore use it as a coarse filter. At each layer, tokens are split into high and low intensity sets based using a dataset-specific threshold. The Top-$K$ tokens are selected by ranking the high-intensity ones via CLS attention scores, with any remainder filled from the low-intensity pool (Fig.~\ref{fig:intensity}). This method is primarily intended to test whether incorporating spectrogram intensities improves attention-only sparsification and how it compares to frequency-aware approaches. \\

\begin{figure}[t]
\centering
  \includegraphics[width = 1.0\linewidth]{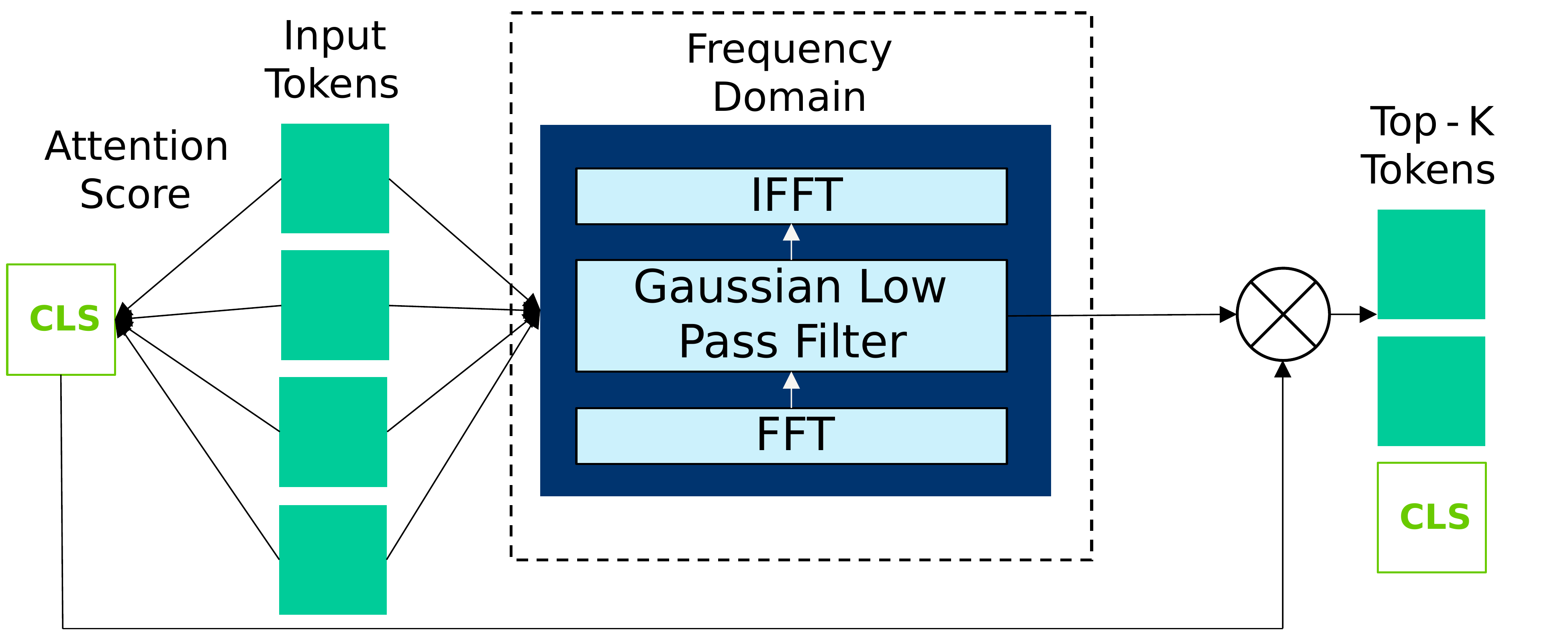}
  \vspace{0.3em}
  \caption{Low-frequency Energy token pruning. Token embeddings are transformed to the frequency domain, where a Gaussian low-pass filter highlights low-frequency structure. The resulting energy scores are fused with attention scores to guide top-K token selection.}
  \label{fig:low_frequency}
  \vspace{-2em}
\end{figure}

\noindent \textbf{Low-Frequency Energy.}  
Inspired by the Low-Frequency Components (LFC) model for vision transformers, we adopt their idea of using low-frequency energy (LFE) for token sparsification \cite{wang2022vtclfc}. In their work, low-pass filtering of token embeddings was shown to preserve task-relevant content more effectively, aligning with the observation that transformers tend to favor low-frequency features \cite{dosovitskiy2021an, wang2022anti}. We extend this idea to audio by computing low-frequency energy within each token embedding, after patch extraction and projection. This provides a local measure of how much of the token's representation is concentrated in low-frequency components. This score is combined with the token’s attention weight, allowing the model to prioritize tokens that are both attended and rich by low-frequency energy, as illustrated in Fig.~\ref{fig:low_frequency}.\\

\begin{figure}[t]
\centering
  \includegraphics[width = 1.0\linewidth]{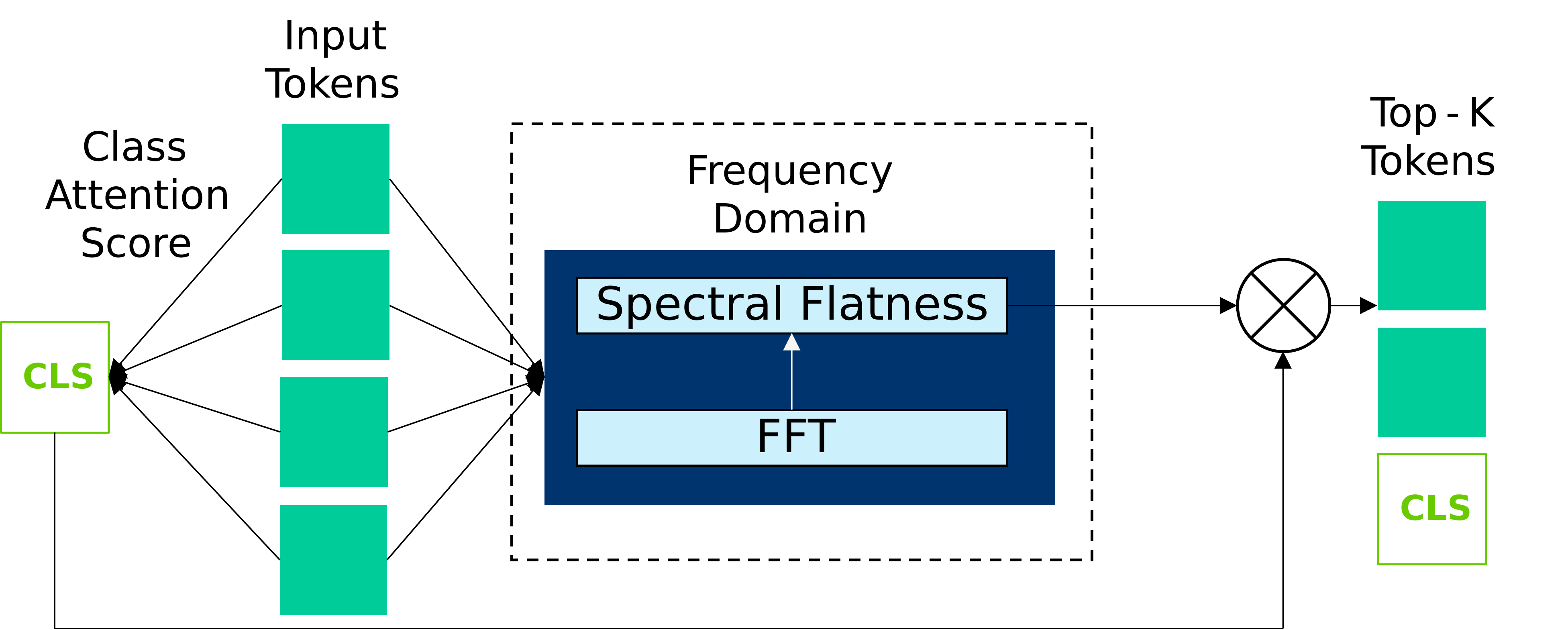}
  \vspace{0.3em}
  \caption{Spectral flatness-guided token pruning. Token embeddings undergo Fast Fourier Transform (FFT) along the embedding dimension to compute spectral flatness, capturing tonal richness. These scores are combined with CLS attention to rank token importance.}
  \label{fig:spectral_flatness}
  % \vspace{-1.5em}
\end{figure}

\begin{figure}[t]
  \centering
  \subfloat[Example Mel spectrogram \cite{speechcommands}.]{
    \includegraphics[width=0.45\linewidth]{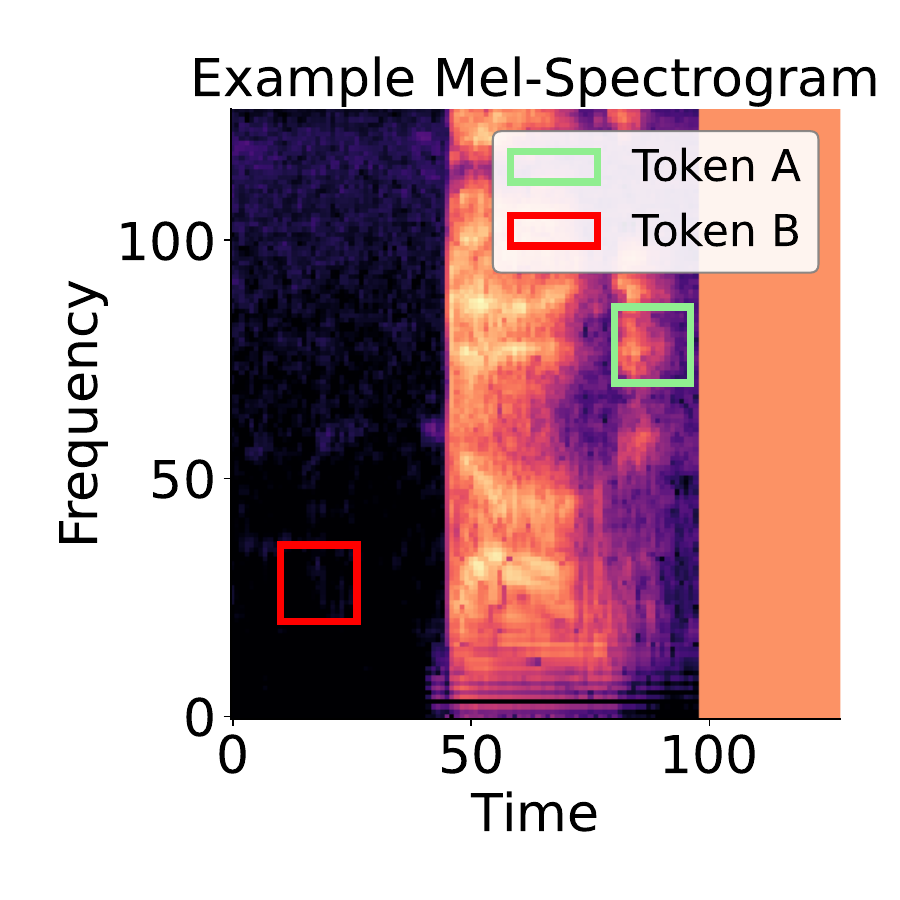}
    \label{fig:spectrogram_sf}
  }\hfill
  \subfloat[Corresponding magnitude spectra.]{
    \includegraphics[width=0.45\linewidth]{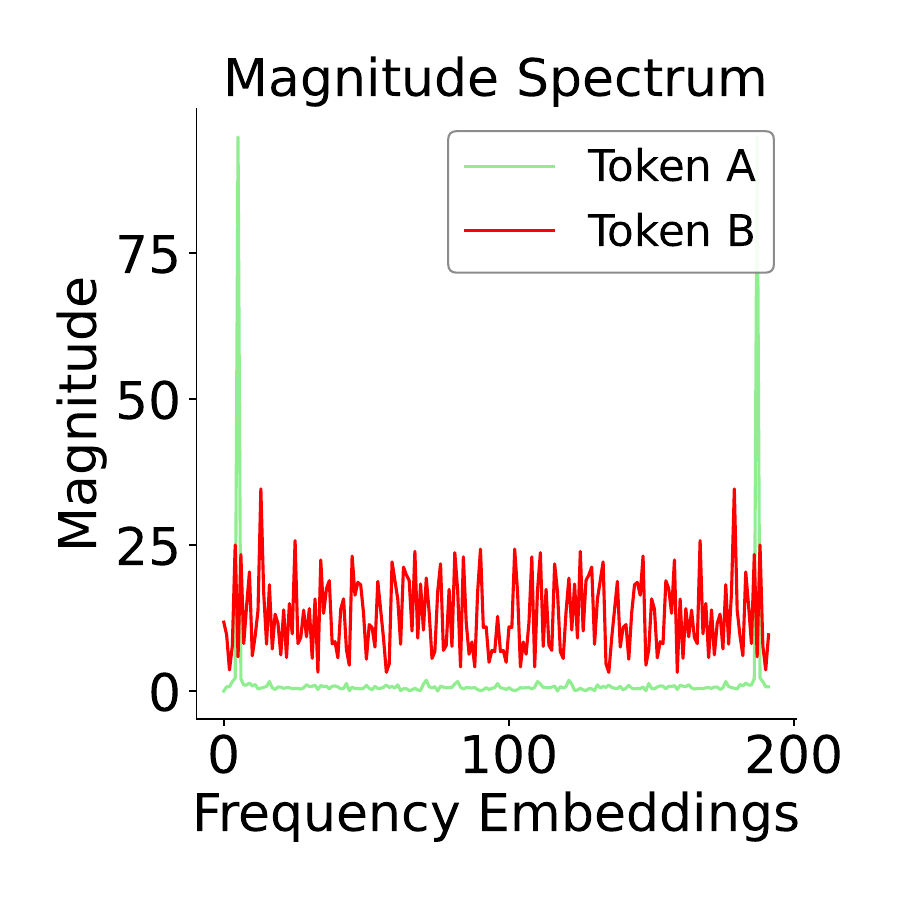}
    \label{fig:spectrum_sf}
  }
  \vspace{0.5em}
  \caption{Spectral flatness at the token level. (a) A spectrogram with two highlighted tokens. (b) FFT of corresponding token embeddings shows Token A has tonal peaks, while Token B has a flatter, noise-like spectrum - making Token A potentially more informative.}
  \label{fig:token_flatness_example}
\end{figure}

\noindent \textbf{Spectral Flatness.} Spectral Flatness (SF) is a widely used measure in audio signal processing for quantifying the degree to which a signal resembles noise or contains structured, tonal components~\cite{jayant1993digital, peeters2004large}. It is computed as the ratio of the geometric mean to the arithmetic mean of a signal’s magnitude spectrum, with lower values indicating pronounced spectral peaks typical of harmonic or speech-like content, and higher values corresponding to flatter, noise-like distributions~\cite{madhu2009note, boashash2015time}. Motivated by these properties, we extend the use of SF to transformer token embeddings by computing it for each token, as illustrated in Fig.~\ref{fig:spectral_flatness}. We treat each embedding as a learned signal, using SF as a proxy for its internal organization, and hypothesize that lower flatness values indicate more structured and content-rich tokens. An illustrative example of this phenomenon is shown in Fig.~\ref{fig:token_flatness_example}, where two spectrogram patches yield token embeddings with contrasting spectral profiles. The token’s flatness score is combined with its attention to the CLS token, prioritizing tokens that are both spectrally structured and highly attended.\\

\vspace{-1.5em}
\subsection{Evaluation Strategy}
\label{sec:evaluation_strategy}

We aim to systematically evaluate token sparsification methods (Sec.~\ref{sec:baselines}) across multiple audio benchmarks, while also investigating the role of input- and frequency-aware token importance estimation strategies (Sec.~\ref{sec:impstrats}). To understand both the practical and performance implications of sparsification, we evaluate all methods under two settings:

\begin{itemize}

    \item \textbf{Post-training Sparsification.} Sparsification is applied only during inference, using already finetuned models.
        \item \textbf{Sparsified Fine-tuning.} Sparsification is applied during fine-tuning, starting from ImageNet-pretrained weights and adapting to the target audio dataset.
\end{itemize}

Using these two settings, we aim to understand the benefits and tradeoffs of sparsified fine-tuning over post-training sparsification, especially at lower keep rates.\\

\section{Experimental Setup}
\label{sec:exp_setup}

\noindent \textbf{Datasets.} For this analysis, three well-established datasets are selected, each covering different audio sample lengths and task complexities. \textbf{Speech Commands V2 (SPC)}~\cite{speechcommands} comprises 105,829 \textbf{1-second} recordings of 35 spoken commands. The dataset is divided into 84,843 training samples, 9,981 validation samples, and 11,005 test samples. \textbf{Environment Sound Classification (ESC50)}~\cite{esc50} consists of 2,000 \textbf{5-second} environmental sound recordings categorized into 50 classes. Its 5-fold cross-validation setup includes 1,600 training samples per fold, ensuring a reliable evaluation framework. \textbf{AudioSet (AS-20K)}~\cite{audioset} is a large-scale multi-class dataset containing \textbf{10-second} audio clips sourced from YouTube, labeled across 527 categories. The balanced training set consists of around 22,000 samples, while the evaluation set includes around 20,000 samples. \\

\noindent \textbf{Backbone.} We adopt the AST (Audio Spectrogram Transformer)~\cite{gong2021} architecture as our backbone for all experiments. The sparsification methods are integrated into the AST model pretrained on ImageNet~\cite{imagenet}, using the standard DeiT-based~\cite{touvron2021deit} configurations: Base (B), Small (S), and Tiny (T), with patch sizes of $16 \times 16$ and an overlap of 6 in both time and frequency, consistent with the default AST recipe.\\

\noindent \textbf{Sparsification Allocation.} As in previous work, the token sparsification module is placed between the attention and multi-layer perceptron (MLP) branches of each transformer block of the AST \cite{Bolya_Fu_Dai_Zhang_Feichtenhofer_Hoffman_2023}. The most common and effective practice is to apply sparsification at three equally spaced points in the transformer, effectively dividing the network into four sections~\cite{kong2022spvit, liang2022evit, long2022beyond}. This is the configuration adopted in our experiments. At each sparsification stage, a fixed keep rate $r \in \{0.25, 0.50, 0.70, 0.90\}$
is applied uniformly across all sparsified layers.\\

% \vspace{1em}

\noindent \textbf{Implementation Details.} All training and fine-tuning procedures follow the original AST setup \cite{gong2021}. Specifically, AS-20K and ESC50 are finetuned for 25 epochs, while SPC is trained for 30 epochs. When token sparsification is enabled, all default masking strategies (e.g., SpecAugment~\cite{park2019specaugment}) in AST are disabled to avoid redundancy, as sparsification inherently performs a form of data masking. All model variants (AST-B, AST-S, AST-T) are evaluated under the post-training sparsification setting to assess performance and efficiency across different model scales. For the sparsified fine-tuning experiments, we use only AST-B, as it is the default model size in the original AST implementation. To quantify the compute cost associated with different sparsification methods, we report the number of Floating Point Operations (FLOPs) per instance using a batch size of 1, measured via the DeepSpeed profiler~\cite{deepspeed}. 
\vspace{-0.5em}

\begin{table*}[t]
\centering
\caption{GFLOPs of AST-B model on AS-20K after sparsification across keep rates (\textbf{\textit{r}}). $\downarrow$ denotes percentage reductions over baseline. Maximum reductions per keep rate are highlighted in \textcolor{green!50!black}{green}.}
\label{tab:flops_audioset}
\footnotesize
\setlength{\tabcolsep}{3pt}
\begin{tabular}{lcccc}
\toprule
& \multicolumn{4}{c}{\textbf{AS-20K (GFLOPs)}} \\
\cmidrule(lr){2-5}
\textbf{Baseline} & \multicolumn{4}{c}{261.43} \\
\cmidrule(lr){2-5}
\textbf{\textit{r} (\%)} & 90 & 70 & 50 & 25 \\
\midrule
Top-K~\cite{haurum2022tokens} &
$222.28^{\textcolor{green!50!black}{\text{\tiny $\downarrow$ 15.0\%}}}$ & 
$164.16^{\textcolor{green!50!black}{\text{\tiny $\downarrow$ 37.2\%}}}$ & 
$124.85^{\textcolor{green!50!black}{\text{\tiny $\downarrow$ 52.3\%}}}$ & 
$93.74^{\textcolor{green!50!black}{\text{\tiny $\downarrow$ 64.3\%}}}$ \\
EViT~\cite{evit2022} &
$222.61^{\text{\tiny $\downarrow$ 14.8\%}}$ & 
$164.45^{\text{\tiny $\downarrow$ 37.1\%}}$ & 
$125.05^{\text{\tiny $\downarrow$ 52.2\%}}$ & 
$93.99^{\text{\tiny $\downarrow$ 64.0\%}}$ \\
ToMe~\cite{Bolya_Fu_Dai_Zhang_Feichtenhofer_Hoffman_2023} &
$226.11^{\text{\tiny $\downarrow$ 13.5\%}}$ & 
$165.77^{\text{\tiny $\downarrow$ 36.6\%}}$ & 
$125.60^{\text{\tiny $\downarrow$ 52.0\%}}$ & 
-- \\
\midrule
Intensity &
$222.28^{\textcolor{green!50!black}{\text{\tiny $\downarrow$ 15.0\%}}}$ & 
$164.16^{\textcolor{green!50!black}{\text{\tiny $\downarrow$ 37.2\%}}}$ & 
$124.89^{\text{\tiny $\downarrow$ 52.2\%}}$ & 
$93.79^{\text{\tiny $\downarrow$ 64.1\%}}$ \\
LFE~\cite{wang2022vtclfc} &
$222.66^{\text{\tiny $\downarrow$ 14.8\%}}$ & 
$164.46^{\text{\tiny $\downarrow$ 37.1\%}}$ & 
$125.10^{\text{\tiny $\downarrow$ 52.1\%}}$ & 
$93.95^{\text{\tiny $\downarrow$ 64.1\%}}$ \\
SF &
$222.41^{\text{\tiny $\downarrow$ 14.9\%}}$ & 
$164.26^{\text{\tiny $\downarrow$ 37.1\%}}$ & 
$124.95^{\text{\tiny $\downarrow$ 52.2\%}}$ & 
$93.84^{\text{\tiny $\downarrow$ 64.1\%}}$ \\
\bottomrule
\end{tabular}
\end{table*}
% \vspace{-1em}

\section{Results \& Analysis}

\subsection{Post-Training Sparsification}
\label{sec:inference_results}

% \subsubsection{Accuracy Across Methods and Model Sizes}
\noindent \textbf{Performance Across Methods and Model Sizes.} Tables~\ref{tab:base-results},~\ref{tab:small-results}, and~\ref{tab:tiny-results} report performance across the AST-B, AST-S, and AST-T models, respectively. Results are shown for all datasets (SPC, ESC50, AS-20K) and across token keep rates from 90\% down to 25\%. We report the average of three runs.

From these results, we find that methods using token frequency information, particularly LFE and SF, consistently perform well across model sizes and datasets, frequently ranking among the top two methods in our setup. ToMe also shows competitive performance, especially on the Small and Tiny models with the ESC50 dataset, but its inconsistency on the AST-B model may suggest sensitivity to model scale or dataset characteristics. We hypothesize that this may be due to ToMe's merging strategy, which can interfere with token semantics more noticeably in larger models where attention patterns are more specialized \cite{kenton2019bert}.\\

\begin{figure}[t]
    \centering
    \fbox{\includegraphics[width=0.95\linewidth]{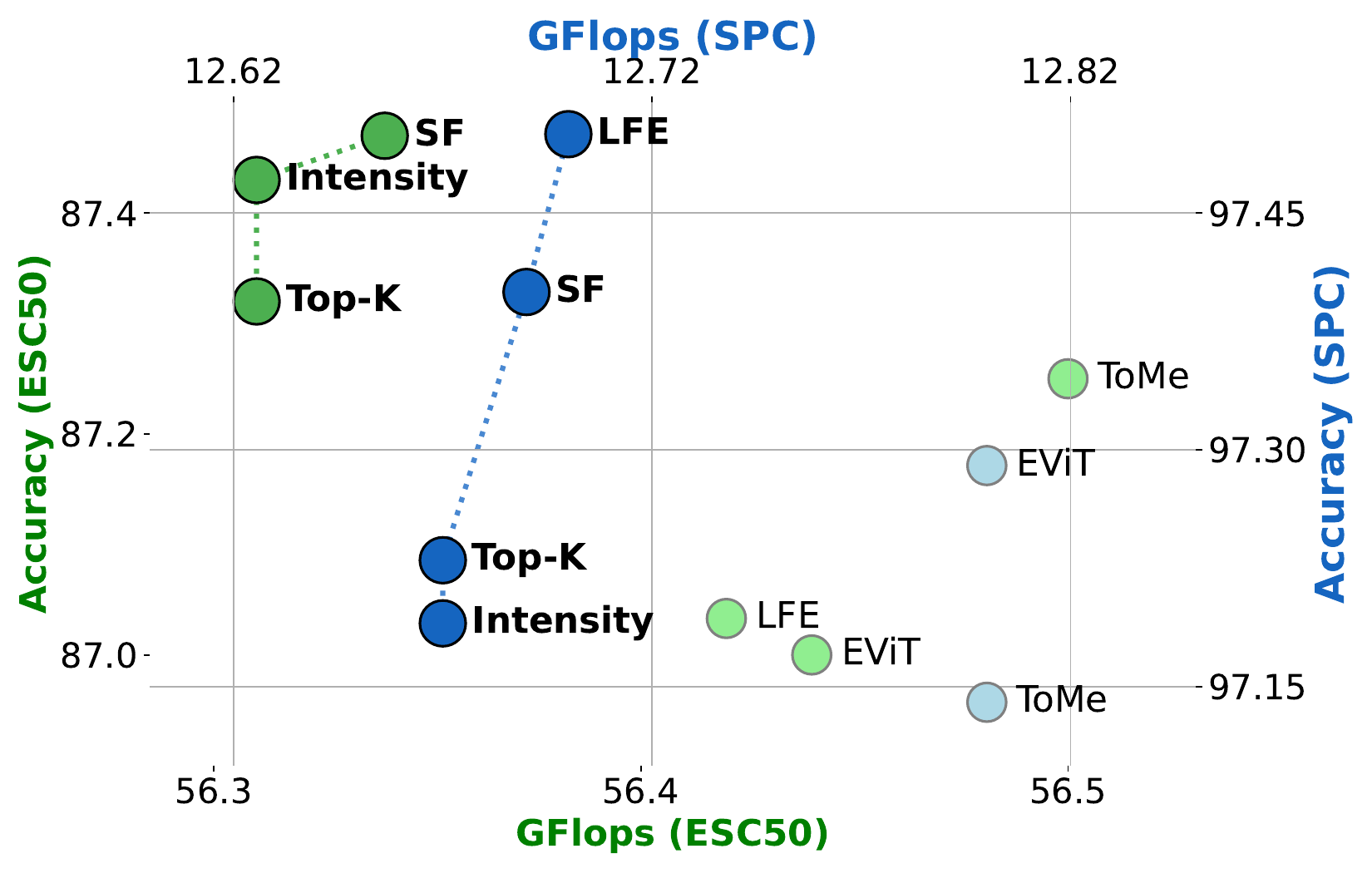}}
    % \fbox{\parbox[b][5cm][c]{8cm}{\centering Figure Placeholder}}
    \vspace{0.5em}
\caption{Post-training sparsification results on the AST-B \cite{gong2021} model evaluated on SPC and ESC50 at a 50\% keep rate. Methods leveraging frequency-domain cues, such as \textbf{Spectral Flatness (SF)} and \textbf{Low-Frequency Energy (LFE) \cite{wang2022vtclfc}}, outperform other methods in terms of accuracy.}
\label{fig:pareto_combined}
% \vspace{-1em}
\end{figure}

\noindent \textbf{Compute Cost Analysis.} We report compute cost in terms of FLOPs for AS-20K using the AST-B model in Table~\ref{tab:flops_audioset}. As expected, the number of FLOPs decreases consistently with lower keep rates across all datasets and methods, reflecting the expected reduction in attention computation as fewer tokens are retained. These trends validate that sparsification mechanisms function as intended, with lower keep rates leading to correspondingly lower compute cost. While this downward trend is consistent across methods, the actual savings differ slightly due to method-specific overheads. 
To illustrate the trade-off between performance and compute cost, we present Pareto front plots. In Fig.~\ref{fig:mainPareto}, we show results for AS-20K, while Fig.~\ref{fig:pareto_combined} shows SPC and ESC50 datasets, on AST-B. We find that frequency-domain methods such as \textbf{Spectral Flatness (SF)} and \textbf{Low-Frequency Energy (LFE)} consistently yield the highest performance across datasets, despite slightly higher computational overhead compared to simpler methods like Top-K. Results follow similar trends, showing the performance scalability of most methods across model sizes and tasks.

\begin{table*}[!t]
\centering
\caption{Post-training sparsification results across model sizes.}
\label{tab:all-results}
\captionsetup[subtable]{justification=centering}
\footnotesize  % or \scriptsize if needed
\setlength{\tabcolsep}{3pt} % tweak to fit

\begin{subtable}[!t]{\textwidth}
% \begin{table*}[t]
\centering
% \vspace{1em}
\caption{Post-training sparsification results for the \textbf{AST-B} model across keep rates (\textit{r}). Best performing methods are in \textbf{bold}, second-best are \underline{underlined}, and results outperforming the AST-B baseline are {\setlength{\fboxsep}{0.25pt}\colorbox{green!30}{\strut highlighted}}.  SF consistently achieves the best performance across most keep rates and datasets, while LFE also performs competitively, particularly on SPC.}
\label{tab:base-results}
% \resizebox{\textwidth}{!}{
\footnotesize
\begin{tabular}{lcccccccccccc}
\toprule
& \multicolumn{4}{c}{\textbf{SPC (Acc(\%))}} & \multicolumn{4}{c}{\textbf{ESC50 (Acc(\%))}} & \multicolumn{4}{c}{\textbf{AS-20K (mAP)}} \\
\cmidrule(lr){2-5} \cmidrule(lr){6-9} \cmidrule(lr){10-13}
\textbf{AST-B} & \multicolumn{4}{c}{97.67} & \multicolumn{4}{c}{89.42} & \multicolumn{4}{c}{0.3572} \\
\cmidrule(lr){2-5} \cmidrule(lr){6-9} \cmidrule(lr){10-13}
\textbf{\textit{r} (\%)} & 90 & 70 & 50 & 25 & 90 & 70 & 50 & 25 & 90 & 70 & 50 & 25 \\
\midrule
Top-K~\cite{haurum2022tokens} &
97.46 & 97.43 & 97.21 & 88.73 &
89.02 & 88.42 & 87.32 & 78.47 &
0.3557 & 0.3546 & 0.3470 & 0.2956 \\
EViT \cite{evit2022} &
97.44 & 97.27 & 97.29 & \underline{90.03} &
88.77 & 88.17 & 87.00 & 75.42 &
0.3552 & 0.3545 & 0.3476 & 0.2981 \\
ToMe~\cite{Bolya_Fu_Dai_Zhang_Feichtenhofer_Hoffman_2023} &
97.51 & 97.47 & 97.14 & -- &
88.76 & 88.45 & 87.25 & -- &
0.3567 & 0.3539 & 0.3470 & -- \\
\midrule
Intensity &
97.55 & 97.46 & 97.19 & 88.63 &
89.10 & \underline{88.52} & \underline{87.43} & 78.62 &
0.3567 & \underline{0.3549} & 0.3479 & \textbf{0.3033} \\
LFE~\cite{wang2022vtclfc} &
\underline{97.66} & \underline{97.62} & \textbf{97.50} & \textbf{93.49} &
\underline{89.12} & 88.45 & 87.02 & \underline{78.92} &
\underline{0.3568} & 0.3546 & \underline{0.3484} & \underline{0.3012} \\
SF &
\textbf{\cellcolor{green!30}97.68} & \textbf{97.66} & \underline{97.40} & 89.00 &
\textbf{89.20} & \textbf{88.67} & \textbf{87.47} & \textbf{78.97} &
\textbf{\cellcolor{green!30}0.3576} & \textbf{0.3564} & \textbf{0.3499} & 0.3007 \\

\bottomrule
\end{tabular}
% }
\end{subtable}
% \end{table*}

\vspace{0.8em}

\begin{subtable}[!t]{\textwidth}
% \begin{table*}[t]
\centering
% \vspace{1em}
\caption{Post-training sparsification results for the \textbf{AST-S} model across keep rates (\textit{r}). Best performing methods are in \textbf{bold}, second-best are \underline{underlined}, and results outperforming the AST-S baseline are {\setlength{\fboxsep}{0.25pt}\colorbox{green!30}{\strut highlighted}}.  SF and LFE dominate across datasets, mirroring the trends observed with AST-B. ToMe performs competitively on ESC50, indicating that token merging may be effective for smaller models.}
\label{tab:small-results}
% \resizebox{\textwidth}{!}{
\footnotesize
\begin{tabular}{lcccccccccccc}
\toprule
& \multicolumn{4}{c}{\textbf{SPC (Acc(\%))}} & \multicolumn{4}{c}{\textbf{ESC50 (Acc(\%))}} & \multicolumn{4}{c}{\textbf{AS-20K (mAP) }} \\
\cmidrule(lr){2-5} \cmidrule(lr){6-9} \cmidrule(lr){10-13}
\textbf{AST-S} & \multicolumn{4}{c}{97.21} & \multicolumn{4}{c}{83.75} & \multicolumn{4}{c}{0.3105} \\
\cmidrule(lr){2-5} \cmidrule(lr){6-9} \cmidrule(lr){10-13}
\textbf{\textit{r} (\%)} & 90 & 70 & 50 & 25 & 90 & 70 & 50 & 25 & 90 & 70 & 50 & 25 \\
\midrule
Top-K~\cite{haurum2022tokens} &
96.97 & 97.04 & 96.83 & 83.60 &
83.10 & 81.65 & 79.23 & 69.03 &
0.3058 & 0.3037 & 0.2995 & 0.2688 \\
EViT~\cite{evit2022} &
96.96 & 97.02 & 96.84 & \underline{89.39} &
83.03 & 81.65 & \underline{79.30} & \underline{69.40} &
0.3052 & 0.3018 & 0.2950 & 0.2672 \\
ToMe~\cite{Bolya_Fu_Dai_Zhang_Feichtenhofer_Hoffman_2023} &
\underline{97.16} & 96.93 & 96.32 & -- &
\textbf{83.33} & \textbf{83.15} & \textbf{80.80} & -- &
0.3057 & 0.3029 & 0.3001 & -- \\
\midrule
Intensity &
96.89 & 96.95 & 96.76 & 83.33 &
83.08 & \underline{82.58} & 78.03 & \textbf{69.60} &
0.3068 & 0.3043 & 0.2996 & 0.2734 \\
LFE~\cite{wang2022vtclfc} &
\textbf{97.21} & \cellcolor{green!30}\underline{97.24} & \textbf{97.21} & \textbf{93.74} &
83.05 & 81.50 & 78.88 & 68.33 &
\underline{0.3089} & \underline{0.3068} & \underline{0.3037} & \underline{0.2792} \\
SF &
\textbf{97.21} & \textbf{\cellcolor{green!30}97.25} & \underline{97.20} & 82.37 &
\underline{83.20} & 81.75 & 79.23 & 69.35 &
\textbf{0.3093} & \textbf{0.3074} & \textbf{0.3049} & \textbf{0.2799} \\
\bottomrule
\end{tabular}
% }
% \end{table*}
\end{subtable}

\vspace{0.8em}

% \begin{table*}[t]
\begin{subtable}[!t]{\textwidth}
\centering
\caption{Post-training sparsification results for the \textbf{AST-T} model across keep rates (\textit{r}). Best performing methods are in \textbf{bold}, second-best are \underline{underlined}, and results outperforming the AST-T baseline are {\setlength{\fboxsep}{0.25pt}\colorbox{green!30}{\strut highlighted}}.  SF and LFE rank quite consistently among the top, showing strong scalability across model sizes, while ToMe excels on ESC50 but is less consistent across other datasets.}
\label{tab:tiny-results}
% \resizebox{\textwidth}{!}{
\footnotesize
\begin{tabular}{lcccccccccccc}
\toprule
& \multicolumn{4}{c}{\textbf{SPC (Acc(\%))}} & \multicolumn{4}{c}{\textbf{ESC50 (Acc(\%))}} & \multicolumn{4}{c}{\textbf{AS-20K (mAP) }} \\
\cmidrule(lr){2-5} \cmidrule(lr){6-9} \cmidrule(lr){10-13}
\textbf{AST-T} & \multicolumn{4}{c}{96.99} & \multicolumn{4}{c}{81.15} & \multicolumn{4}{c}{0.2759} \\
\cmidrule(lr){2-5} \cmidrule(lr){6-9} \cmidrule(lr){10-13}
\textbf{\textit{r} (\%)} & 90 & 70 & 50 & 25 & 90 & 70 & 50 & 25 & 90 & 70 & 50 & 25 \\
\midrule
Top-K~\cite{haurum2022tokens} &
96.37 & 96.21 & 95.93 & 88.06 &
80.58 & 78.65 & 73.95 & 56.23 &
0.2721 & 0.2699 & 0.2601 & 0.2189 \\
EViT~\cite{evit2022} &
96.35 & 96.19 & 95.87 & \textbf{92.37} &
80.35 & 78.53 & 73.78 & \textbf{57.03} &
0.2668 & 0.2632 & 0.2591 & 0.2161 \\
ToMe~\cite{Bolya_Fu_Dai_Zhang_Feichtenhofer_Hoffman_2023} &
\underline{97.01} & 96.26 & 95.00 & -- &
\textbf{\cellcolor{green!30}81.25} & \textbf{79.53} & \underline{74.13} & -- &
\textbf{0.2731} & \underline{0.2701} & 0.2603 & -- \\
\midrule
Intensity &
\textbf{\cellcolor{green!30}97.02} & 96.62 & 96.12 & 89.12 &
80.47 & \underline{79.01} & 73.61 & 54.91 &
0.2695 & 0.2654 & 0.2599 & 0.2233 \\
LFE~\cite{wang2022vtclfc} &
97.00 & \textbf{\cellcolor{green!30}97.03} & \textbf{96.81} & \underline{91.36} &
80.40 & 78.70 & 73.45 & 53.68 &
0.2699 & 0.2675 & \underline{0.2624} & \underline{0.2391} \\
SF &
\textbf{\cellcolor{green!30}97.02} & \underline{96.98} & \underline{96.71} & 87.87 &
\underline{80.63} & 78.70 & \textbf{74.20} & \underline{56.73} &
\underline{0.2728} & \textbf{0.2709} & \textbf{0.2638} & \textbf{0.2399} \\
\bottomrule
\end{tabular}
% }
\end{subtable}
\end{table*}

\subsection{Sparsified Fine-tuning}

\begin{figure*}[t]
  \centering
  \includegraphics[width=0.6\textwidth]{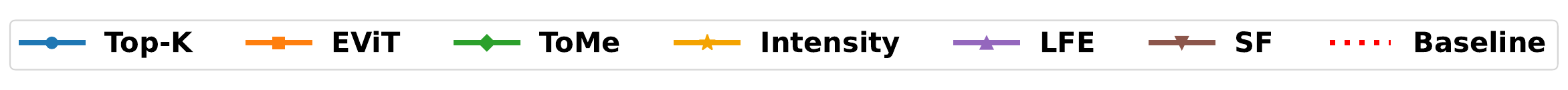}\\[0.3em]
  \subfloat[SPC]{
    \fbox{\includegraphics[width=0.29\textwidth]{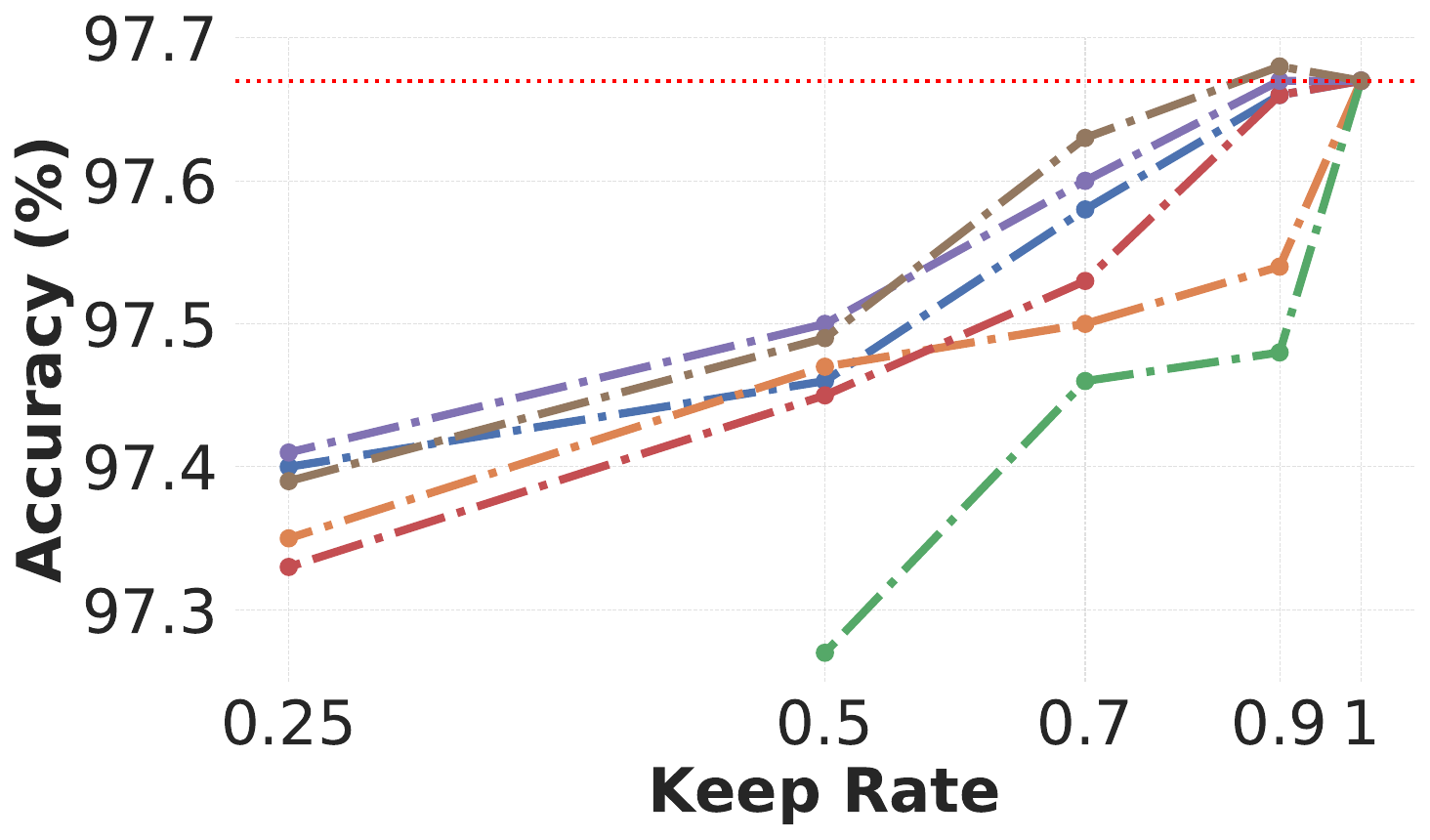}}
    \label{fig:finetuning_spc}
  }\hfill
  \subfloat[ESC50]{
    \fbox{\includegraphics[width=0.29\textwidth]{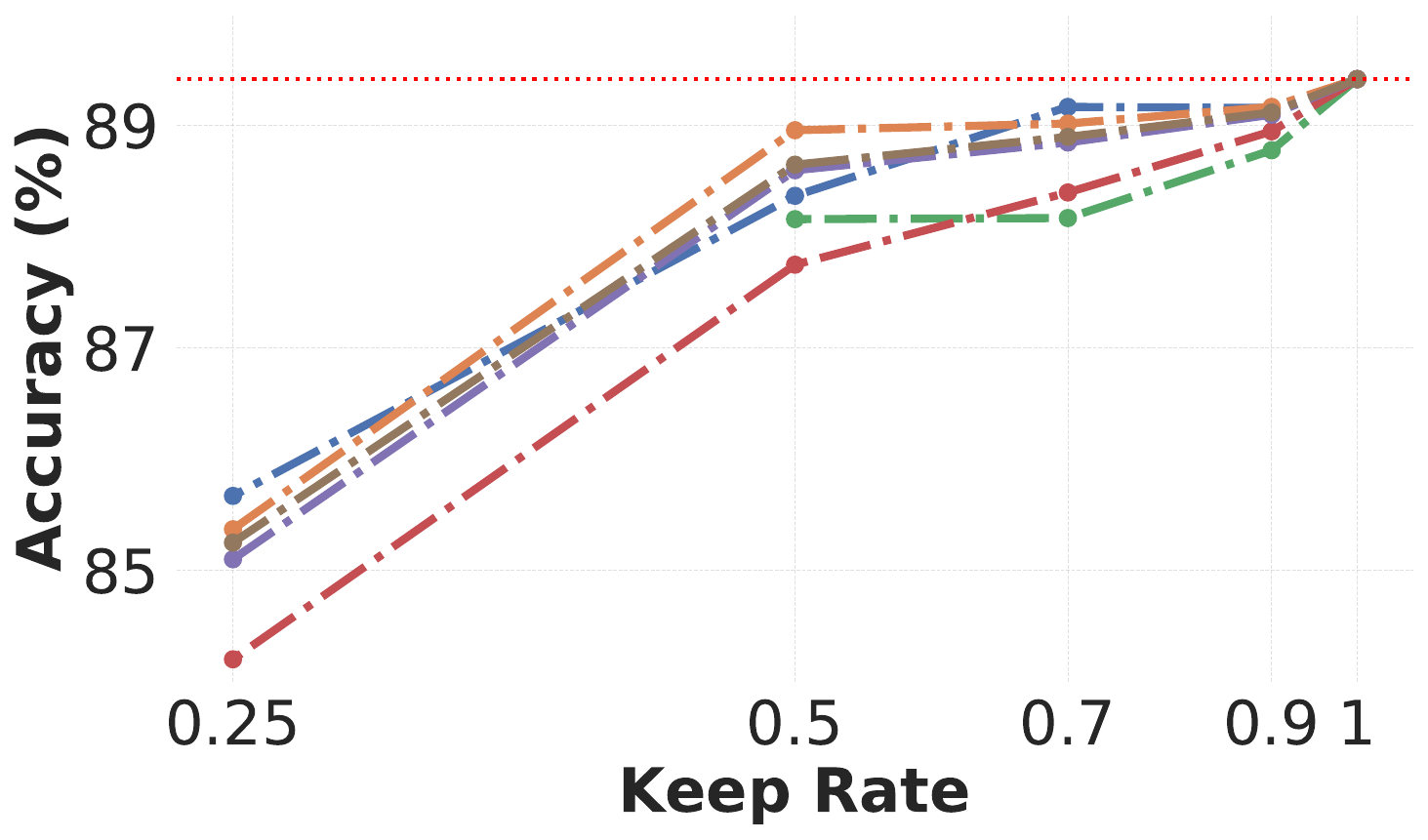}}
    \label{fig:finetuning_esc50}
  }\hfill
  \subfloat[AS-20K]{
    \fbox{\includegraphics[width=0.29\textwidth]{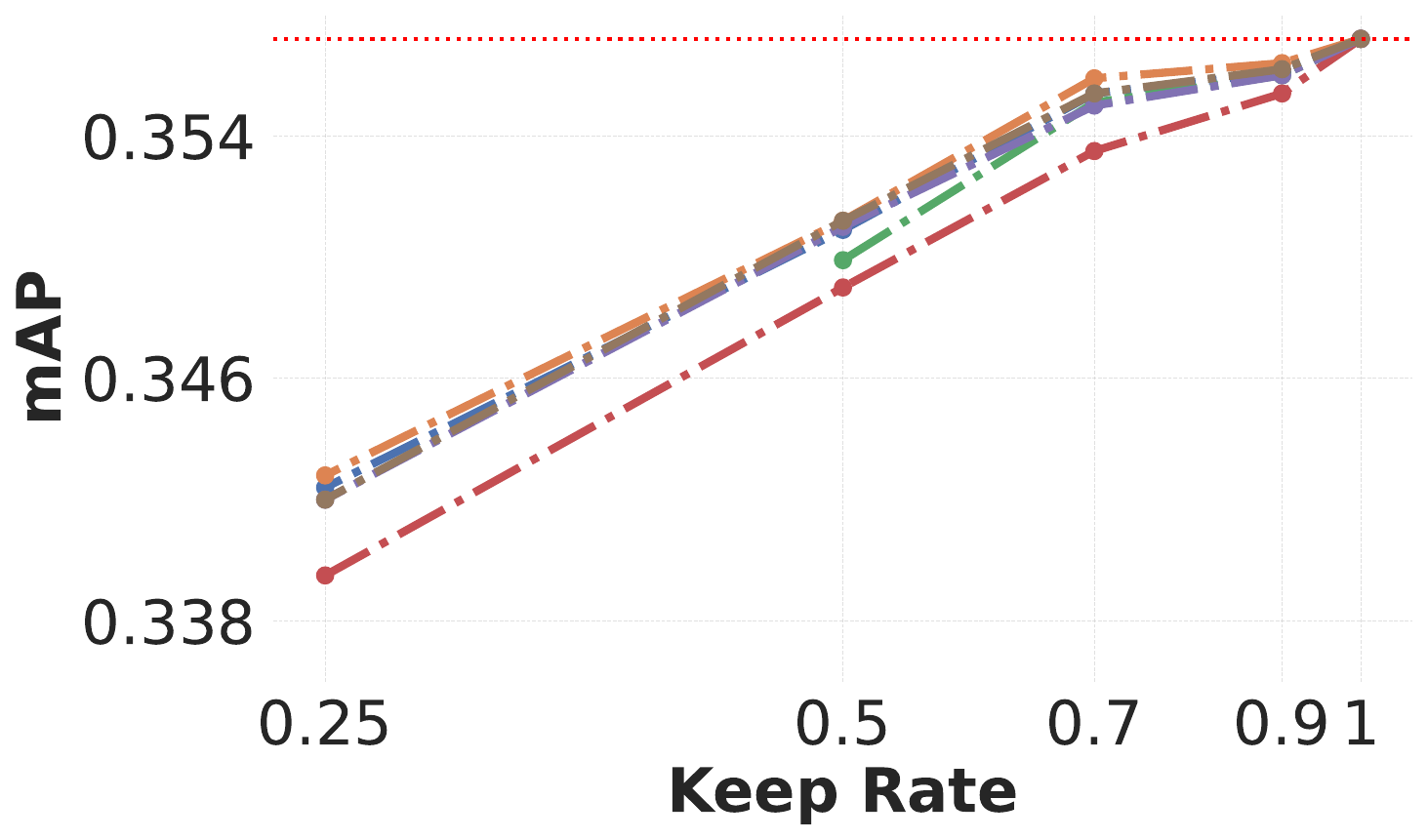}}
    \label{fig:finetuning_as20k}
  }
  \caption{\textbf{Sparsified Fine-tuning:} Performance across varying keep rates. Frequency-based methods (SF, LFE), as well as pure attention-based pruning methods (Top-K, EViT), consistently maintain high performance across tasks. Most methods retain competitive performance down to a 50\% keep rate. Overall, the performance differences between methods are not very significant, indicating that multiple sparsification methods can be equally effective when fine-tuned.}
  \label{fig:finetuning}
\end{figure*}

Fig.~\ref{fig:finetuning} shows how task performance varies with decreasing keep rates across datasets under sparsified fine-tuning, with results averaged over three runs. 

Across all datasets, we observe that performance remains robust even under aggressive sparsification, with most methods maintaining accuracy close to the baseline at a 25\% keep rate. This indicates that when sparsification is integrated during training, the model is able to adapt to the imposed sparsity, learning to preserve task-relevant information despite reduced number of tokens. Furthermore, the performance differences between methods are less pronounced than in the post-training setting.  This suggests that the model learns to accommodate whichever tokens are retained during training, thereby reducing sensitivity to the specific scoring heuristic. As a result, even simple methods like Top-K remain competitive, and the relative advantage of more informed strategies, such as SF and LFE, is reduced.

In summary, sparsified fine-tuning makes the model less sensitive to the specific sparsification heuristic, as it learns to adapt to the retained tokens. This contrasts with post-training sparsification, where token importance is fixed and the choice of heuristic plays a more decisive role.

\begin{figure*}[t]
  \centering
  \includegraphics[width=0.5\textwidth]{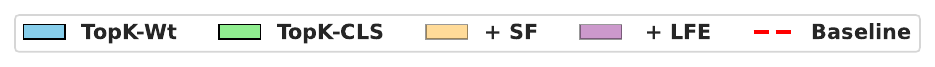}\\[0.3em]
  \subfloat[SPC]{
    \includegraphics[width=0.31\textwidth]{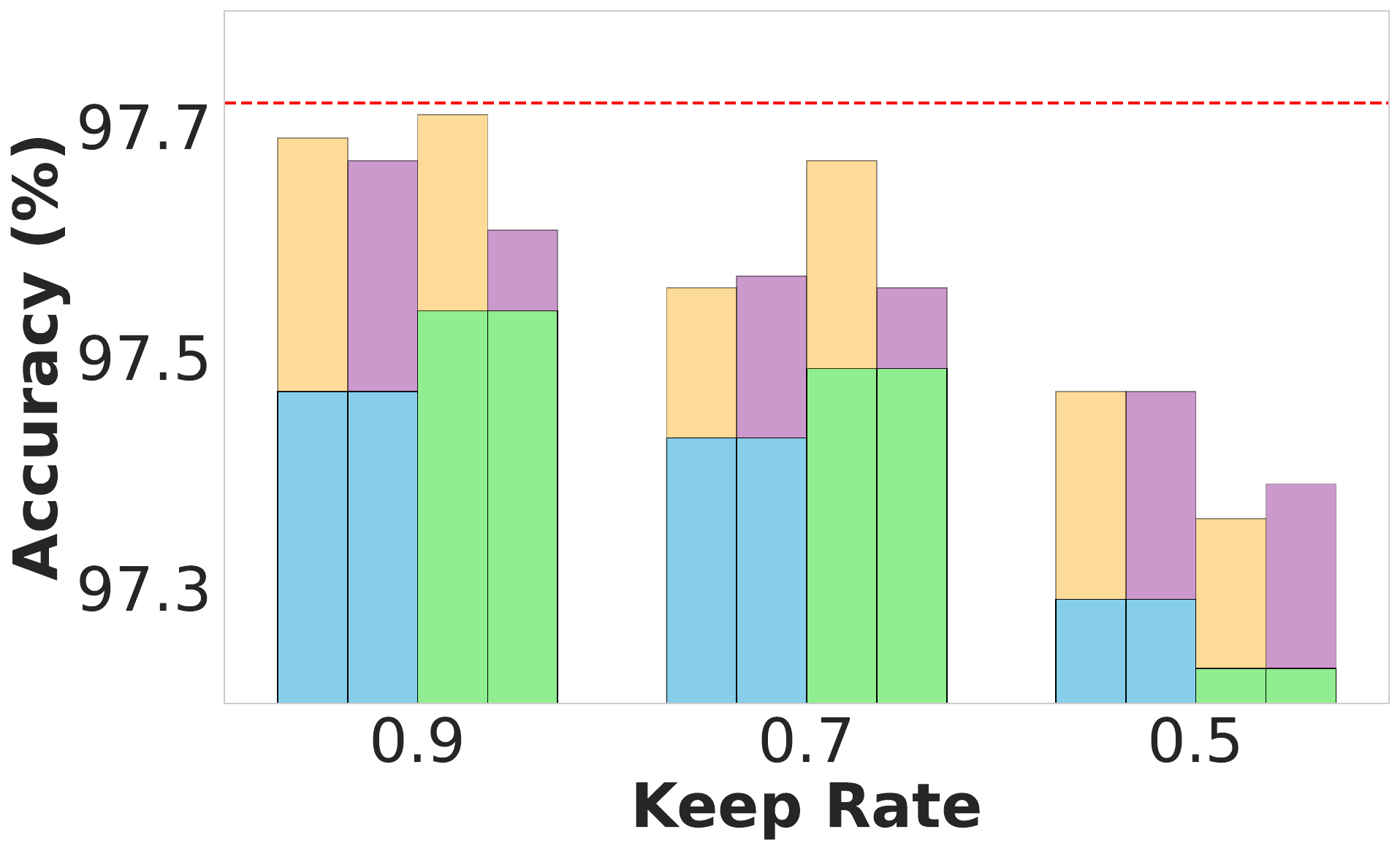}
    \label{fig:spc_frqPlot}
  }\hfill
  \subfloat[ESC50]{
    \includegraphics[width=0.31\textwidth]{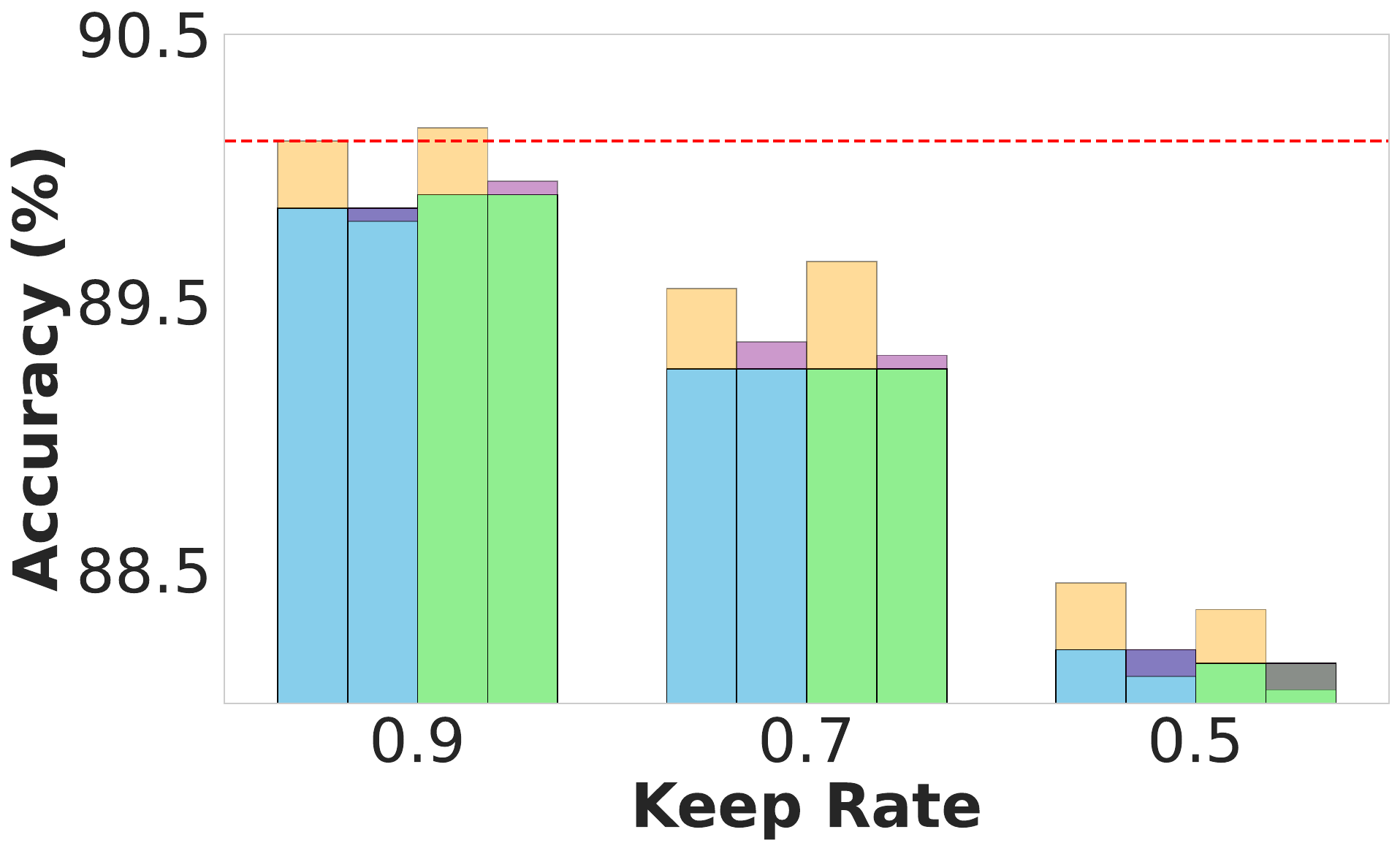}
    \label{fig:esc50_frqPlot}
  }\hfill
  \subfloat[AS-20K]{
    \includegraphics[width=0.31\textwidth]{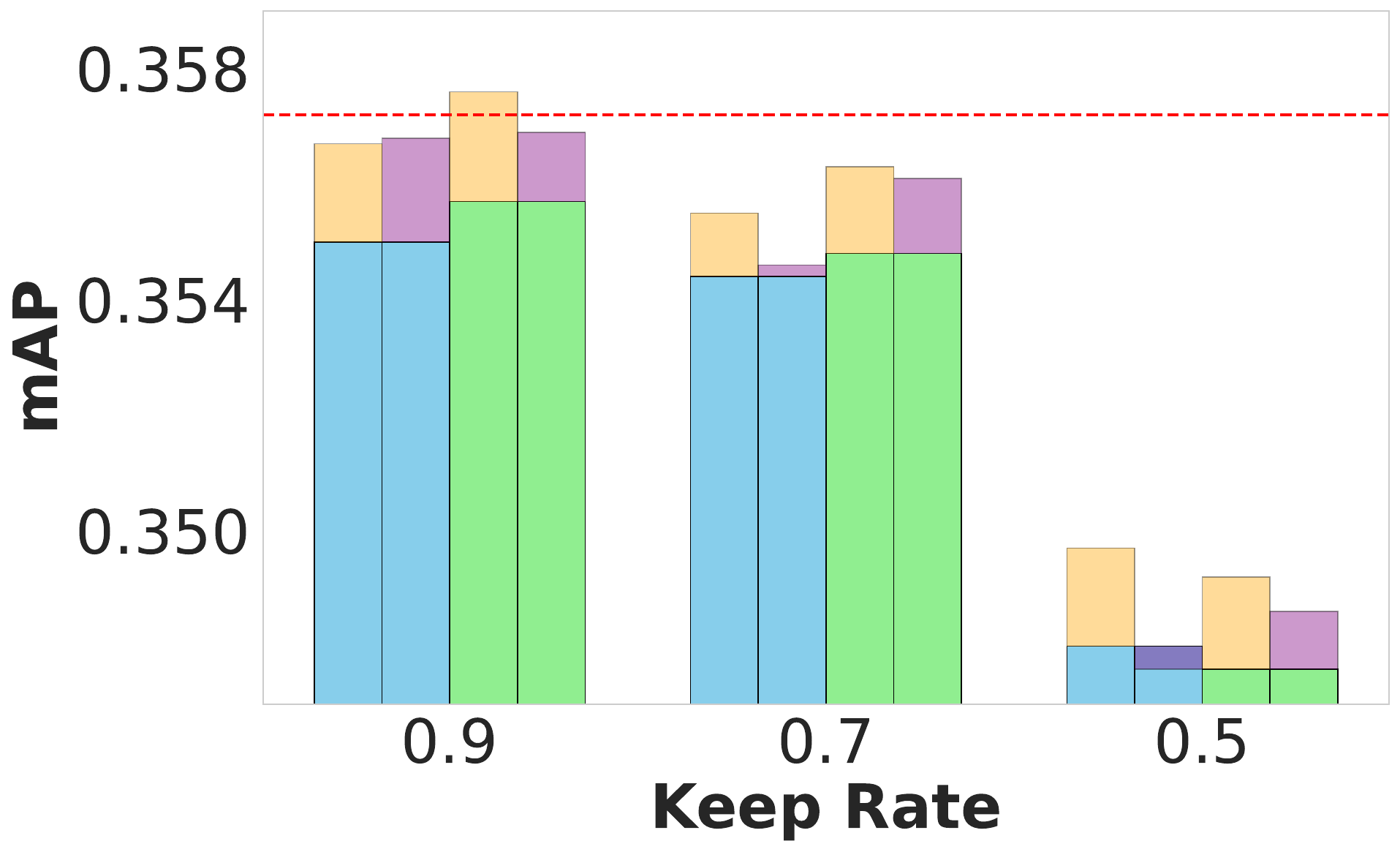}
    \label{fig:as20k_frqPlot}
  }
  \caption{Performance comparison of frequency-based methods (LFE and SF) against attention-only baselines (TopK-Wt and TopK-CLS). SF slightly but consistently outperforms attention-based baselines across keep rates, while LFE shows competitive results but occasionally underperforms on ESC50 and AS-20K.}
  \label{fig:frq_improvements}
  \vspace{-1em}
\end{figure*}

\subsection{Comparative \& Controlled Studies}
\label{sec:ablation}
\noindent \textbf{Contribution of Frequency-domain Information.}  
To understand the source of performance differences between SF and LFE, we conduct a comparative study that isolates the effects of their frequency metrics and attention formulations. SF uses standard CLS-token attention (TopK-CLS), and LFE on the other hand, uses a head-weighted attention variant (TopK-Wt)~\cite{wang2022vtclfc}. To isolate the impact of these factors, we evaluate all four combinations of frequency metric (SF or LFE) and attention type (TopK-CLS or TopK-Wt). Each configuration is run five times to account for randomness.

As shown in Fig.~\ref{fig:frq_improvements}, SF consistently outperforms its TopK-only counterparts across keep rates and attention variants. In contrast, while LFE generally improves over its baseline, it occasionally underperforms on AS-20K and ESC50, suggesting that emphasizing low-frequency components may not universally benefit performance and may depend on dataset characteristics. Furthermore, TopK-Wt tends to outperform TopK-CLS, especially at lower keep rates ($<70\%$), which may be because TopK-Wt also includes peer-token attention, offering a more distributed importance signal than CLS-token attention alone. 

Overall, incorporating frequency-domain information yields modest but consistent gains over attention-only pruning methods in post-training settings. 

\vspace{1em}

\noindent \textbf{Effect of Input Duration on Intensity-Based Pruning.} In Section \ref{sec:inference_results} (Table \ref{tab:base-results}), we observed that intensity-based pruning outperforms purely attention-based methods (Top-K, EViT) on ESC50 and AS-20K across most settings, whereas its advantage is less pronounced on SPC. This relatively strong performance on datasets with longer audio clips (5 and 10 seconds, respectively) - compared to weaker results on SPC (with 1-second clips) raises the question of a potential dependency on input duration. 

To investigate this, we conduct a controlled experiment comparing intensity-based pruning with a purely attention-based Top-K method. We apply random temporal windowing at inference time on AS-20K and ESC50, simulating input lengths ranging from 1 second to the full duration of each dataset (i.e., 1s to 10s for AS-20K, and 1s to 5s for ESC50). Each result is averaged over six runs to account for randomness. The sparsification configuration is fixed to the most aggressive setting, with a keep rate of only 25\%.

\begin{figure}[t]
  \centering
  \includegraphics[width=0.7\linewidth]{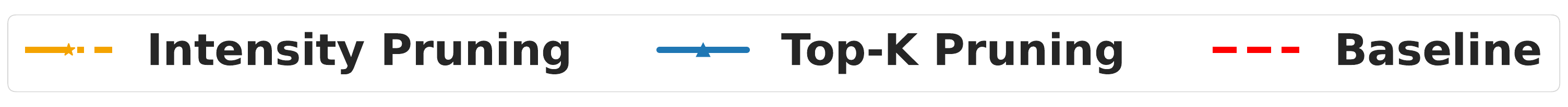}\\[0.3em]
  \subfloat[AS-20K]{
    \includegraphics[width=0.45\linewidth]{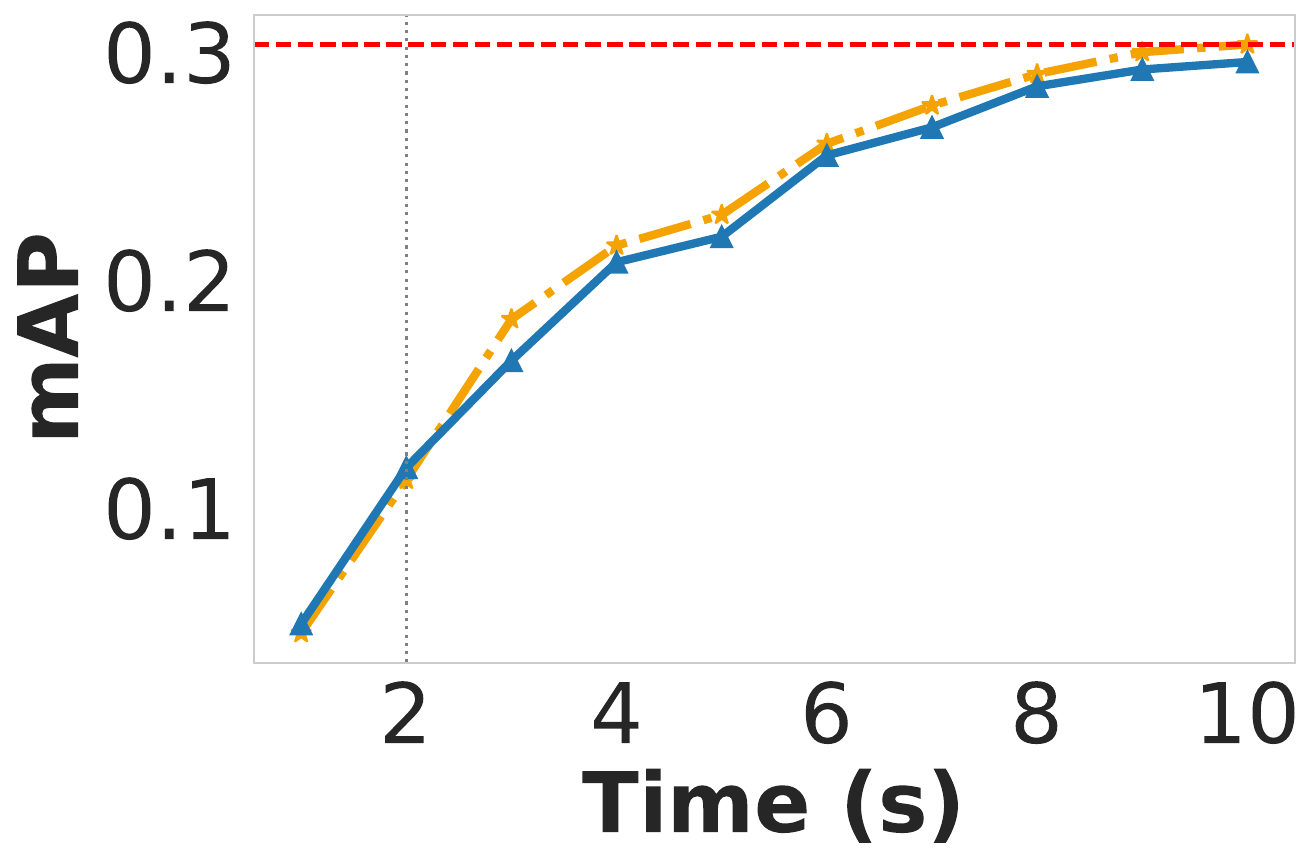}
    \label{fig:audioset_window}
  }\hfill
  \subfloat[ESC50]{
    \includegraphics[width=0.45\linewidth]{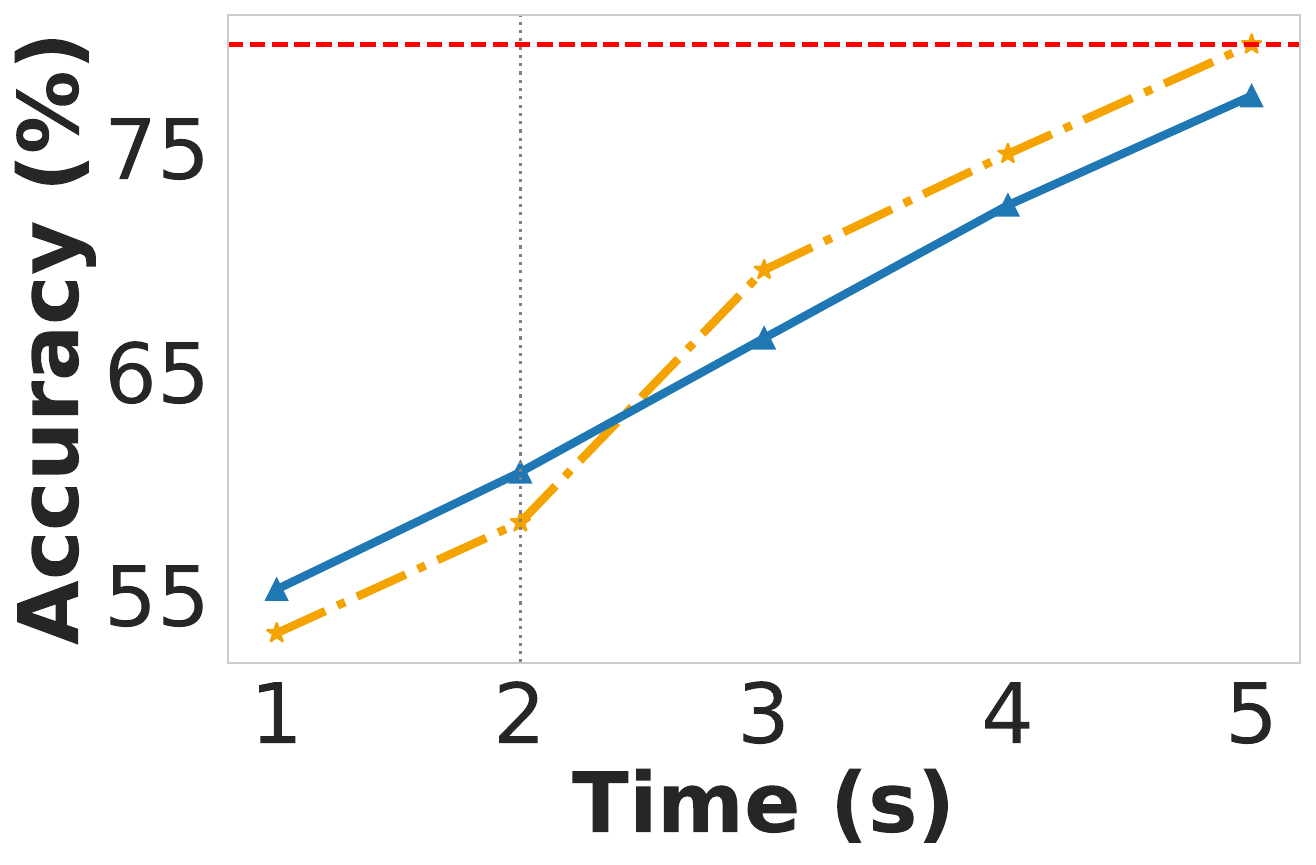}
    \label{fig:esc50_window}
  }
  \caption{Token pruning performance as a function of input duration. Intensity Pruning consistently outperforms Top-K pruning for longer audio windows ($>2$ seconds), suggesting that spectrogram intensity is a more informative proxy for longer audio samples.}
  \label{fig:window_analysis}
  % \vspace{-0.3em}
\end{figure}

As shown in Fig.~\ref{fig:window_analysis}, intensity-based pruning consistently outperforms the Top-K method for audio segments longer than 2 seconds across both datasets. These results provide further empirical support for our hypothesis and suggest that intensity-based pruning works better for longer inputs. We attribute this effect to the interaction between input duration and token redundancy. First, longer audio clips yield more tokens and often contain repeated or sustained events, introducing redundancy. Therefore, even after sparsification, enough informative content remains. Second, they tend to include clear low-intensity regions, such as silence or background noise, which can be safely discarded. In contrast, short clips (1-2 seconds) typically contain a single spoken word or sound event, where nearly all tokens contribute to  the core signal. These inputs offer little redundancy and minimal variation in intensity, making it harder to distinguish unimportant tokens. As a result, sparsification based on intensity can mistakenly remove critical information, leading to performance degradation.

\subsection{Practical Guidelines}
Based on our analysis, we outline several recommendations for effectively applying token sparsification in audio transformers:
\begin{itemize}
    \item Frequency-aware heuristics, particularly Spectral Flatness (SF) and Low-Frequency Energy (LFE), quite consistently outperform other methods. SF stands out as one of the most practical choices for post-training sparsification, as it requires no additional tuning and generalizes well across configurations. LFE, while also effective, involves tuning a low-pass filter cutoff to maximize performance, which may require dataset-specific calibration. 
    \item For scenarios involving longer input durations (e.g., $>$ 2 seconds), intensity-based pruning can also outperform other standard methods mainly for larger model sizes, but it requires threshold tuning to distinguish between high- and low-energy patches. 
    \item While token merging methods like ToMe can perform well on small models or specific tasks, their effectiveness is more variable and less consistent across certain tasks.
    \item Finally, when applying sparsification during fine-tuning, the specific heuristic becomes less critical, as the model can adapt to sparsity - enabling even simple methods to perform well. 
\end{itemize}

\vspace{2em}

\section{Conclusion}

In this work, we presented a comprehensive evaluation of token sparsification techniques for audio transformers, covering pruning and merging approaches across multiple model sizes, datasets, and deployment scenarios. Our results show that frequency-aware methods, particularly those drawing on spectral flatness and low-frequency energy, consistently outperform attention-only methods, particularly in post-training sparsification settings. We also introduced a novel strategy that combines attention with spectral priors, which proved effective across a wide range of configurations. These findings underscore the value of integrating audio-specific cues into sparsification strategies, moving beyond generic attention signals. Looking forward, we see opportunities to explore more advanced task-specific token estimation strategies and extend these techniques to real-time applications.

\newpage
\bibliographystyle{preprint2col}  
\bibliography{references}

\end{document}